# Realizing an Isotropically Coercive Magnetic Layer for Memristive Applications by Analogy to Dry Friction

M. Mansueto[1], A. Chavent[1], S. Auffret[1], I. Joumard[1], J. Nath[1], I. M. Miron[1], U. Ebels[1], R. C. Sousa[1], L. D. Buda-Prejbeanu[1], I. L. Prejbeanu[1] and B. Dieny[1]

[1] Univ. Grenoble Alpes, CEA, CNRS, Grenoble INP, IRIG-SPINTEC, 38000 Grenoble, France

**Abstract**

We investigate the possibility of realizing a spintronic memristive device based on the dependence of the tunnel conductance on the relative angle between the magnetization of the two magnetic electrodes in in-plane magnetized tunnel junctions. For this, it is necessary to design a free layer whose magnetization can be stabilized along several or even any in-plane direction between the parallel and the antiparallel magnetic configurations. We experimentally show that this can be achieved by exploiting antiferromagnet-ferromagnet exchange interactions in a regime where the antiferromagnet is thin enough to induce enhanced coercivity and no exchange bias. The frustration of exchange interactions at the interfaces due to competing ferro- and antiferromagnetic interactions is at the origin of an isotropic dissipation mechanism yielding isotropic coercivity. From a modeling point of view, it is shown that this isotropic dissipation can be described by a dry friction term in the Landau-Lifshitz-Gilbert equation. The influence of this dry friction term on the magnetization dynamics of an in-plane magnetized layer submitted to a rotating in-plane field is investigated both analytically and numerically. The possibility to control the free layer magnetization orientation in an in-plane magnetized magnetic tunnel junction by using the spin transfer torque from an additional perpendicular polarizer is also investigated through macrospin simulation. It is shown that the memristor function can be achieved by the injection of current pulses through the stack in the presence of an in-plane static field transverse to the reference layer magnetization, the aim of which is to limit the magnetization rotation between 0° and 180°.

## I. INTRODUCTION

The interest in neuromorphic computing has increased considerably in the last decade because of its efficiency in applications related to big data analysis. Indeed, the promises offered by artificial intelligence (AI) are so outstanding that AI is now being used in many applications, from consumer electronic applications (in smartphone and computers) to applications in the health sector (assistance in medical diagnostics), in future autonomous cars, in industrial and domestic robots, in security (recognition systems, cybersecurity), in military and space fields, and so on. So far, all the existing implementations use conventional processor hardware to perform AI processing. In AI, very large amounts of data have to be processed. With conventional processor architecture, the corresponding power consumption associated with transferring the data between memory and logic blocks is quite high. In this context, it would be very power efficient to develop specific neural processor units formed of highly interconnected arrays of neurons and memristive synapses [1,2]. Memristors can be designed with some adaptation from the technology of magnetic non-volatile memory [3]. In the last years, several ideas have been investigated. A first concept based on domain wall-based magnetic tunnel junctions (MTJs) was published in 2016 [4]. The resistance variation is achieved through displacement of a domain wall (DW) in the free layer of a perpendicular MTJ. In this way, the ratio between the parallel and antiparallel states of the MTJ is defined by the position of the DW whose displacement is controlled through spin transfer torque (STT) induced by current pulses. The reliability of the device depends on the possibility to control the pinning of the DW, which is not easy to achieve. Moreover, the need to stabilize a high number of intermediate resistance states requires a device of a large size, which leads to a scalability problem. In a second approach, spin orbit torque (SOT) was exploited in an antiferromagnet which was used both as a heavy metal line and to provide exchange bias to a perpendicularly magnetized ferromagnet [5]. In this three-terminal device, the SOT of the bilayer is large enough to switch the magnetization and the use of an in-plane field can be avoided thanks to the presence of the induced exchange bias field. In this device, the proportion of reversed magnetization can be controlled by current pulses allowing an analog resistance variation. However, the three-terminal nature of the device and related lateral sizes make its downsizing scalability difficult. A third proposed idea is based on the statistical STT switching of several MTJs connected in series [6]. The use of field and current pulses of different amplitudes allows a good control of $N + 1$ resistance states, where N is the number of MTJs. For achieving more states, the devices must be reset at each step. In addition, several concepts of memristive devices involving nonmagnetic materials have been proposed, which are derived from other technologies of nonvolatile memories such as resistive oxide random access memory involving ion diffusion in oxides or phase change RAM based on controlled transition between amorphous and crystalline states [7]. In most of those cases, challenges concern reliability, variability, and endurance. In this context, we propose a spintronic memristive device that based on the recently developed magnetic random-access memories (MRAM) technology, can offer several advantages. The idea is to achieve several intermediate resistance states in a single nanopillar in-plane magnetized MTJ by exploiting the angular dependence of the tunneling magnetoresistance (TMR). Indeed, it is known that the conductance of a MTJ varies with the cosine of the angle between the two magnetizations [free layer (FL) and analyzer (A)] reaching, respectively, its maximum and minimum values in parallel (P) and antiparallel (AP) configurations [8]. The stabilization of these intermediate magnetization angles would give rise to the desired memristive characteristic. To control the in-plane rotation of the free layer magnetization, an out-of-plane polarizer (POL) is added to the structure [see Fig. 1(a)]. Under dc current [Fig. 1(b)], the STT from such a perpendicular polarizer is known to induce a steady out-of-plane large angle precession of the free layer magnetization [9,10] [Fig. 1(c)]. If instead of a dc current, pulses of current [Fig. 1(d)] are sent through the nanopillar with a duration corresponding to a fraction of the precession period, then the magnetization rotates step by step at each pulse [Fig. 1(e)]. Depending on the current pulse polarity, the magnetization rotates clockwise or anticlockwise. However, to obtain a monotonous variation of the resistance under pulses of constant polarity, we need to limit the angular excursion of the free layer magnetization between 0° (parallel configuration with analyzer) and 180° (antiparallel configuration) or less. To achieve this, an in-plane magnetic field transverse to the magnetization of the analyzer A is applied [Fig. 1(f)].

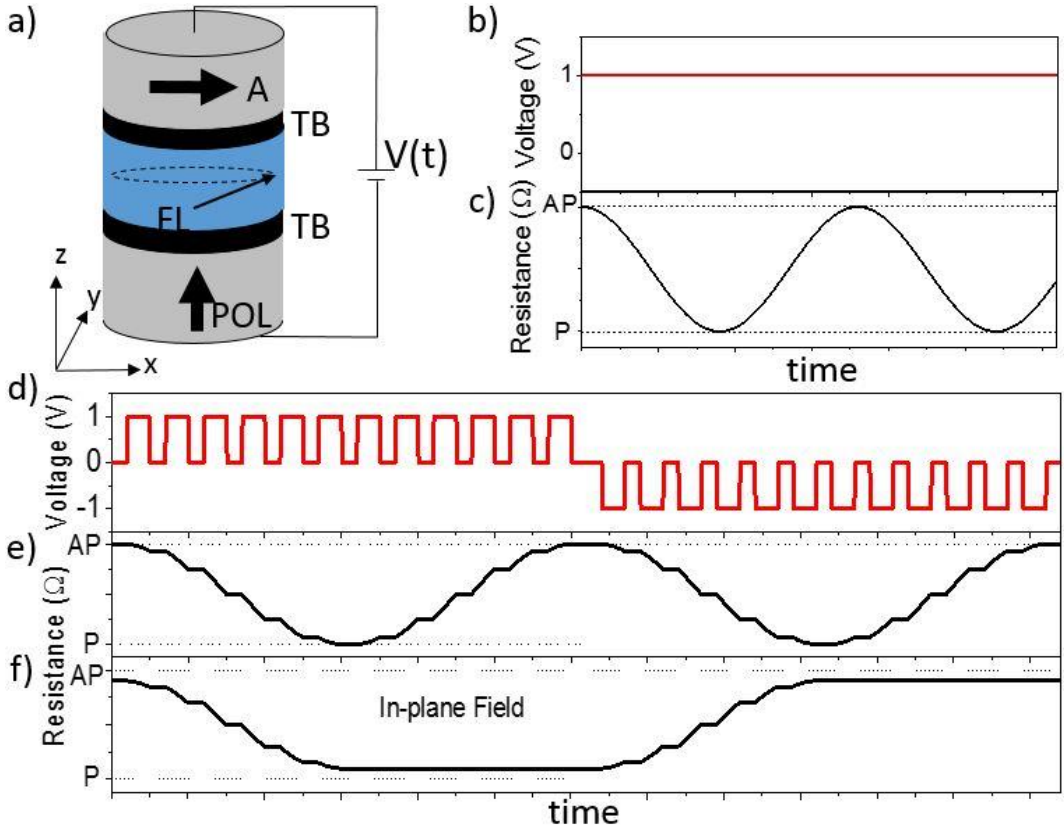

Fig. 1 a) Schematics of the device under DC voltage for which out-of-plane precession occurs. b) DC voltage signal and c) expected resistance evolution. d) Pulsed voltage signal and expected resistance evolution without e) and with f) transverse in-plane field.

The realization of this device requires the development of a key element: an isotropically coercive free layer allowing to stabilize the free layer magnetization in any direction in-plane. In the first part of this paper, we show analytically and through numerical simulations (subsection 1A) that such an isotropic coercivity can be realized if the free layer magnetization is experiencing a so-called dry friction, which consists, as in mechanics, of a constant torque opposite to the magnetization motion. Such a dry friction was already reported in a number of experimental situations including random anisotropy systems [11,12]. In subsection 1B, we demonstrate that a free layer with dry friction can be experimentally realized with a ferromagnet (F)-antiferromagnet (AF)-ferromagnet (F) sandwich where the thickness of the antiferromagnet is low enough not to yield exchange bias but is large enough to yield a maximum in coercivity. In the second part of the paper based on numerical simulations, predictions on the effect of STT on the system are made with and without the presence of a static field in order to find the appropriate conditions for achieving a memristive behavior. Finally, we numerically demonstrate that the complete device can effectively behave as a memristor.

## II. PART 1: FIELD-DRIVEN EQUILIBRIUM AND DYNAMICS UNDER DRY FRICTION

### A. Model

In classical mechanics, the concepts of viscous and dry frictions are frequently used in the general equation of motion. According to Newton's law, the position r of a body of mass m under the action of a force f follows the equation

$$m\ddot{\mathbf{r}} = \mathbf{f} - \alpha \dot{\mathbf{r}} - \beta \frac{\dot{\mathbf{r}}}{|\dot{\mathbf{r}}|} \qquad (1)$$

where α and β are positive damping constants. The viscous friction is a term proportional to the velocity of the body $\dot{r}$. The dry friction term, independent of the velocity and against the motion, is often used to describe the friction between solids. Similar to mechanics, the dynamics of magnetization M in ferromagnets also involves various mechanisms of energy dissipation. The Landau-LifshitzGilbert equation (LLG) [13] governing the magnetization dynamics is usually written as

$$\frac{\partial \mathbf{M}}{\partial t} = -\gamma \left( \mathbf{M} \times \mu_0 \mathbf{H}_{eff} \right) + \frac{\alpha_G}{M_S} \left( \mathbf{M} \times \frac{\partial \mathbf{M}}{\partial t} \right) \qquad (2)$$

The first term on the right-hand side describes the precessional motion of the magnetization around the effective field Heff. Here, μ0 is the vacuum permeability, γ is the gyromagnetic ratio, and MS is the spontaneous magnetization of the ferromagnet. The second term is the Gilbert dissipative term that is dependent on the "magnetization velocity" as a viscous friction in mechanics. The LLG Eq. (2) is extensively used to perform micromagnetic simulations [14]. In some cases, an extra source of dissipation can be added to the system by introducing specific defects such as pinning sites or distributions of anisotropy or of local magnetization. Another approach consists of introducing in the LLG equation a velocity-independent damping that, as in mechanics, accounts for the presence of distributed defects (for instance, associated with a distribution of anisotropy axes or of exchange interactions). The LLG equation then contains an extra dry friction term

$$\frac{\partial \mathbf{M}}{\partial t} = -\gamma \left( \mathbf{M} \times \mu_0 \mathbf{H}_{eff} \right) + \frac{\alpha_G}{M_S} \left( \mathbf{M} \times \frac{\partial \mathbf{M}}{\partial t} \right) + \beta M_S \frac{\mathbf{M} \times \frac{\partial \mathbf{M}}{\partial t}}{\left| \mathbf{M} \times \frac{\partial \mathbf{M}}{\partial t} \right|} \qquad (3)$$

In the frame of this modified LLG equation, Kittel and Galt [15] and Malozemoff and Slonczewski [16] studied the pinning of domain walls during their motion in continuous thin films. With similar purposes, Baltensperger and Helman in 1991 [17] used the additional dry damping to explain the phenomenon of hysteresis, linked to dissipation but not through the magnetization velocity. This model allowed studying the influence of magnetic friction on the linewidth of ferromagnetic resonance (FMR) [18]. More recently the model has been implemented in micromagnetic simulations to study the field and current driven domain wall motion in nanostrips with defects [19-21]. In the usual LLG (eq.2), the magnetization stops moving when the magnetization vector $\mathbf{M}$ gets aligned with the effective field $\mathbf{H}_{eff}$ since in this case, there is no more torque acting on the magnetization $|\gamma(\mathbf{M} \times \mu_0 \mathbf{H}_{eff})| = 0$. In contrast, the additional dry friction term in eq. 3 generalizes this static equilibrium condition to:

$$|\gamma(\mathbf{M} \times \mu_0 \mathbf{H}_{eff})| < \beta M_S \qquad (4)$$

The dynamics of the magnetization itself is affected being described by a dynamic effective damping $a = \alpha_G + \frac{\beta}{v}$ with $v = \frac{1}{M}\left|\frac{\partial \mathbf{M}}{\partial t}\right| = \frac{1}{M_S^2}\left|\mathbf{M} \times \frac{\partial \mathbf{M}}{\partial t}\right|$.

Hereafter, we used the modified LLG (eq.3) including a dry friction term to analyze the magnetization dynamics of an isotropic magnetic thin film under either a static or an in-plane rotating magnetic field. The analytical predictions are compared with the results obtained by numerical integration of the LLG equation.

## 1. Static equilibrium under in-plane field

The system considered in this study is a ferromagnetic continuous thin film laying in the x-y plane and of thickness $l$ along the $z$-axis. The magnetization is supposed to be uniform and subject only to the demagnetizing field and a static in-plane applied field. The effective field in the LLG (eq.3) is simply given by:

$$H_{eff} = (H_{app}\cos(\varphi_H) \quad H_{app}\sin(\varphi_H) \quad -M_S\cos(\theta_M))$$

where ($\theta_{M,H}$ and $\varphi_{M,H}$ are respectively the polar angle and the azimuthal angle of the magnetization (M) and the external field (H)). Considering an in-plane initial magnetization ($\theta_M = \pi/2$) and a motion within the x-y plane, the equilibrium condition (eq. 4) reduces to:

$$|\sin(\varphi_H - \varphi_M)| < \frac{\beta}{\gamma\mu_0 H_{app}} \qquad (5)$$

Both the applied field amplitude and the dry friction coefficient define the limit angle of a sector in which the magnetization can be stable at equilibrium. Without dry friction, the magnetization is perfectly aligned with the applied field as expected. In contrast, with dry friction, two regimes are identified with respect to a threshold field value $H_{TH} = \beta/(\gamma\mu_0)$, as shown in Fig. 2a (where $\phi = \varphi_H - \varphi_M$) in which, for each value of the applied field, the limit angle of this sector of stability of the magnetization is plotted. For low applied fields $H_{app} < H_{TH}$, the amplitude of the torque acting on the magnetization is not large enough to initiate the motion (mathematically the value of the right-hand term of eq. 5 is larger than 1). In this case, the magnetization is stable along any in-plane direction (Fig. 2b left). At the critical field $H_{TH}$, the torque created by the field is balanced by the friction within the whole half plane around the field direction (white region in Fig. 2b center for which the right-hand term of eq. 5 is equal to one). This *sector of stability* then reduces with increasing the field amplitude (Fig. 2b right for which the right-hand term of eq. 5 is lower than one).

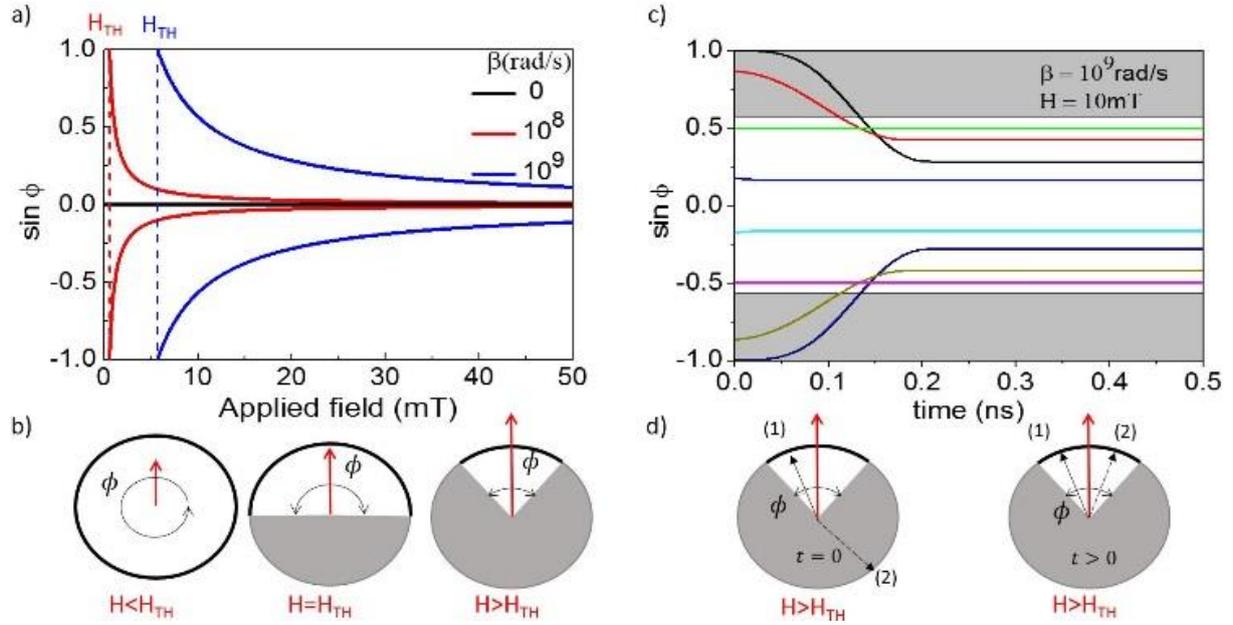

Fig. 2 a) Analytical limit of the sector of stability under in-plane static field for several values of dry friction coefficient $\beta$. b) Schematics of the evolution of the stability region (in white) with field amplitude. c) Macrospin time evolution of magnetization with different initial angles under static magnetic field. The white region corresponds to the analytical sector of stability. d) Schematics of initial (left) and final (right) position of magnetization for initially stable and unstable cases (respectively 1 and 2). The parameters used for all the simulation presented in this work are: $M_S = 10^6 A/m$, $\alpha_G = 0.04$, $K_1 = 0 J/m^3$, demagnetizing tensor $N = (0\ 0\ 1)$.

In Fig. 2c, the time evolution of magnetization for different initial angles is shown for an applied field of 10mT and $\beta = 10^9 rad/s$. Inside the stability sector (white region in Fig. 2c), whatever the direction of the magnetization, eq. 5 is valid and no motion of the magnetization can occur. This means that if the initial magnetization is already inside the stability sector (whose amplitude depends on the applied field as shown by eq. 5) it will stay stable (as green, blue, cyan and magenta cases in Fig. 2c and case (1) in Fig. 2d). In the opposite case, when the initial magnetization is outside the stability sector (in grey in Fig. 2c and d), the magnetization will feel a torque higher than the friction that will initiate a relaxation towards the limits of the cone as for the other cases in Fig. 2c and case (2) in Fig. 2d. The exact final position depends on the relative initial magnetization angle with respect to the field angle. The larger the initial angle with respect to the field direction, the larger the torque, the higher the initial angular velocity of the magnetization and the closer the final position of the magnetization will be to the field. The macrospin simulations confirm that the stability limit between the torque of a static field and the dry friction is given by eq. 5.

Finally note that the effect described in this paragraph does not depend on the absolute initial direction of the magnetization. In fact, all images in Fig. 2 are represented as a function of the difference between the field angle and the magnetization angle to emphasize that the results are invariant under rotation of magnetization and field in the plane. In particular, despite the absence of any anisotropic term in the different energy terms acting of the magnetization, the threshold field due to the dry friction results in a coercive field exhibiting isotropic characteristics in the xy plane. For any in-plane initial magnetization direction, the same field amplitude in the opposite direction is needed to induce magnetization motion.

## 2. Magnetization dynamics under in-plane rotating field

A second case of interest consists in applying an in-plane rotating field of amplitude $H_{rot}$, angular velocity $\omega_{rot}$ and initial direction at t=0 given by $\varphi_{rot}$. This field can induce an in-plane rotation of the magnetization provided the torque due to the field is larger than the dry friction torque. The effective field from eq. 3 becomes:

$$H_{eff} = (H_{rot}cos(\omega_{rot}t + \varphi_{rot}) \quad H_{rot}sin(\omega_{rot}t + \varphi_{rot}) \quad -M_S cos(\theta_M))$$

In the absence of dry friction, the magnetization would follow the direction of the rotating field, after a transient regime, with a small delay dependent on the values of the Gilbert damping, the field amplitude and the frequency of the rotating field. The effect of the dry friction term shows up, similarly to the previous case, by a significant increase of the threshold field $H_{rotTH} = \frac{\beta}{(\gamma\mu_0)} + \frac{\alpha_G \omega_{rot}}{(\gamma\mu_0)}$ separating two regimes: the low field regime where the torque is not sufficient to overcome the friction, and the high field regime for which a stationary rotation of the magnetization is induced. In the first case, the torque induced by the rotating field is balanced by the dry friction torque resulting in a stable magnetization state in any position in the plane. In the second case, for fields higher than a threshold, the magnetization, after a transient regime, starts to follow the rotating field (Fig. 3a) with a drag angle different from zero (Fig. 3b). In this dynamical stationary condition, where $\varphi_H = \omega_{rot}t + \varphi_{rot}$ and $\varphi_M = \omega_{rot}t + \varphi_{rot} - \phi$, the final drag angle can be written :

$$\sin(\varphi_H - \varphi_M) = \sin(\phi) = \frac{\beta}{\gamma\mu_0 H_{rot}} + \frac{\alpha_G \omega_{rot}}{\gamma\mu_0 H_{rot}}. \qquad (6)$$

As in the previously discussed static case, this angle is linearly dependent on the dry friction parameter $\beta$ and inversely proportional to the amplitude of the rotating field. To this, a correction is added due to the Gilbert damping that is proportional to the dynamical parameters $\omega_{rot}$ and $\alpha_G$.

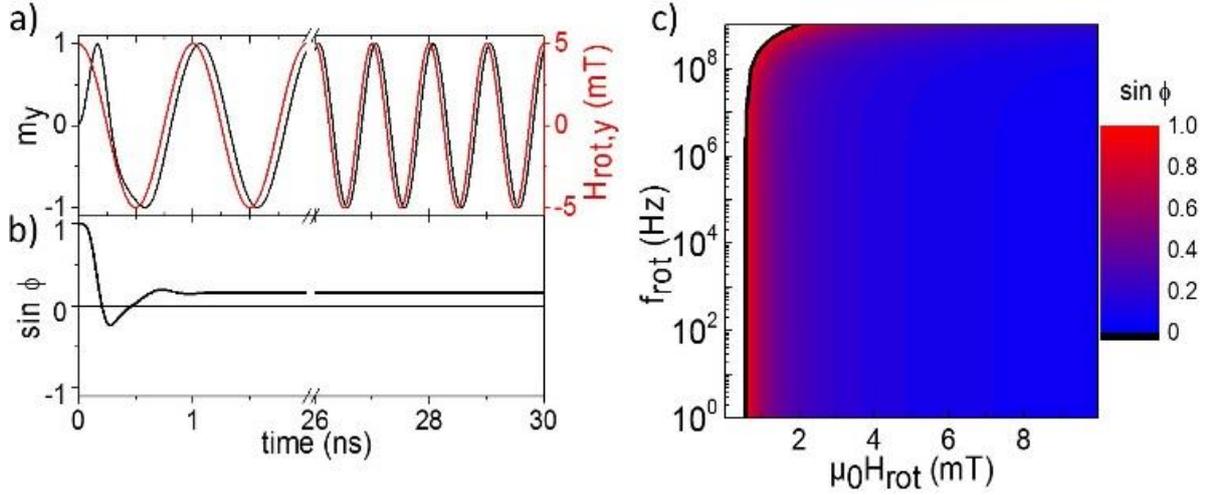

Fig. 3 a) Time evolution of rotating field of 5mT at 1GHz and the response of an in-plane component of the magnetization for $\beta = 10^8 \, rad/s$. b) Time evolution of the drag angle between field and magnetization. c) Field amplitude / field frequency mapping of the stationary angle between field and magnetization. The black line is the analytical expression of the threshold field.

The two cases treated above (static and rotating field) clearly show how the dry friction term in the LLG equation affects both the static and dynamic behavior of the magnetization. In the next paragraph, we show that such dry friction can be realized in ferromagnetic/antiferromagnetic/ferromagnetic sandwiches and analyze their dynamic magnetic behavior in light of the results presented above.

**B. Experiments**

Early studies in 1987 [11] and 1993 [22] reported the experimental and numerical observations of dry friction effects in ferromagnetic systems with distributed axes of anisotropy. In such systems (e.g., amorphous GdxDy(1−x)Ni rare-earth/transition metal alloys [11]), the dry friction arises from coupled spins or grains that, having an isotropically distributed anisotropy direction, tend to fall into their potential minimum during a collective motion of the magnetization. The associated dissipation is enhanced for a proper ratio between the random anisotropy and the exchange energy. Here, we propose an alternative system with similar dry friction-like behavior, which can be integrated in MTJs. The idea is to exploit the frustration of exchange interactions, which exist at the interface between a F and an AF due to competing positive and negative exchange interactions, similar to an interfacial spin glass [23]. In the past, several publications have discussed the influence of the antiferromagnet thickness on the exchange bias and coercive field of such F-AF bilayers [24,25]. After annealing under a magnetic field, an exchange bias appears above a certain thickness of the AF layer (typically above 2 nm for IrMn at room temperature (RT) [26]). This results from the fact that above this thickness, the anisotropy of the AF layer becomes sufficiently large for the AF spin lattice to resist the interfacial torque exerted by the F magnetization on the AF spin lattice upon field cycling. However, below this critical AF thickness, the AF spin lattice is fully dragged due to the torque caused by the F magnetization, yielding dissipation (coercivity) but no exchange bias. A maximum in coercivity is observed for an AF thickness corresponding to this critical thickness. For these low AF thicknesses, the interfacial frustration makes the AF spin lattice so disordered that it exhibits spin glasslike isotropic properties. Its dragging upon the motion of the F magnetization is expected to yield a dissipation equivalent to a dry friction, as in random anisotropy systems [11].

We performed an experimental study to investigate the AF thickness dependence of coercivity and exchange bias field in un-patterned Py(1nm)/IrMn($l_{IrMn}$,)/Py(1nm) trilayers, deposited by sputtering (Py=Permalloy=$Ni_{80}Fe_{20}$). These samples were annealed at 300°C for 1h30mins under an in-plane field of 0.23T.

Fig. 4a shows the influence of the IrMn thickness ($l_{IrMn}$) on the exchange bias and coercive field in these trilayer systems. All experiments described in the paper were performed at room temperature. As expected, for a critical thickness of IrMn (2.1nm here), an enhanced coercivity and a zero exchange bias are measured with vibrating sample magnetometer technique (VSM). This critical thickness of IrMn was therefore selected in the subsequent experimental studies.

### 1. Rotational hysteresis

Hysteresis loop measurements are performed on the sample with the VSM technique. As shown in Fig. 4(b), a first hysteresis loop is measured with the field applied parallel to the annealing field (corresponding to $\phi H = 0$). Then other measurements are performed with the field applied in different directions characterized by the in-plane angle $\phi H$. For each field direction, a single loop is observed indicating that the two ferromagnetic layers are strongly coupled through the thin AF layer. Moreover, the variations of the coercive field and of the remanence between loops at different in-plane angles are below 5%. Therefore, as expected, such a F-AF-F sandwich can be considered as exhibiting an isotropic coercivity (similar to the model described in Sec. II A 1). Because the AF spin lattice is fully dragged upon field cycling, the AF layer exerts a dry friction on the F magnetization independent of the direction of application of the field.

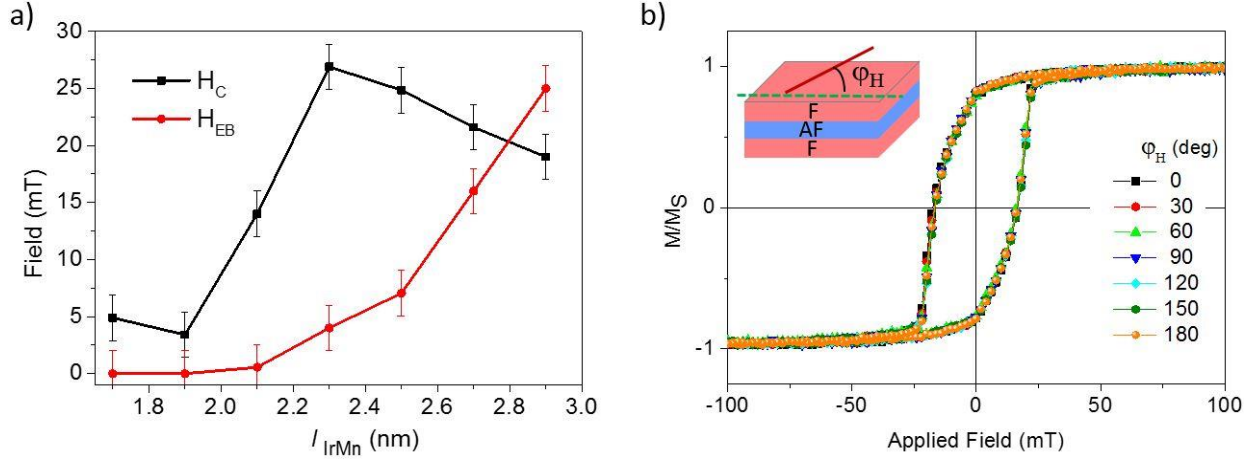

Fig. 4 a) Coercive field ($H_C$) and exchange bias ($H_{EB}$) dependence on AF thickness obtained with VSM technique. b) Field loops at different in-plane angle $\varphi_H$ measured with VSM technique for $l_{IrMn}$ = 2.1nm. The angle is defined as shown in the inset, where the dashed green line is the annealing field direction and the red one the applied field direction.

**2. Planar Hall effect measurements**

Planar Hall Effect (PHE) [27] measurements are performed on the Py/IrMn/Py trilayer under a rotating field, similar to the simulations described in Sec. 2 related to the influence of a rotating field on the free layer magnetization dynamics in the presence of dry friction. A rotating magnetic field of various amplitudes (0–34 mT) and varying frequencies up to 10 Hz is used. The sample is connected to the voltage and current terminals in a Hall geometry allowing to measure the planar Hall resistance (RPHE). The angular dependence of this parameter is described as

$$R_{PHE} = \frac{V_H}{I} = \frac{\Delta R}{2}\sin(2\varphi_M)$$

where VH is the Hall voltage, R is the PHE magnetoresistance, and $\phi_M$ is the in-plane angle of the magnetization with respect to the current direction [28]. When dry friction comes into play, if the rotating field amplitude is large enough, the magnetization is expected to follow the rotating field with a certain drag angle φ = ϕ$_H$ − ϕ$_M$ given by Eq. (6) (where ϕ$_H$ is the angle of the field with respect to the current direction). This should result in a PHE signal varying as

$$R_{PHE} = \frac{V_H}{I} = \frac{\Delta R}{2}\sin(2\varphi_H - 2\phi)$$

This formula can be applied for extracting the value of the drag angle φ for different values of the field amplitude, and thereby for deriving the β parameter from Eq. (6). The experimental results, shown in Figs. 5(a) and 5(b), are obtained by first saturating the in-plane magnetization along the current direction and then applying the rotating field of the selected amplitude. As expected, for field amplitudes lower than the coercive field [black line in Fig. 5(a)], the magnetization is not able to follow the field. When the field becomes higher than the threshold value, following an initial transient regime, a sin(2ϕ$_H$) dependence of the PHE signal is observed with a phase shift dependent on the field amplitude. This is consistent with the general picture that the magnetization is rotating with the field with a drag angle due to dry friction. The fact that the amplitude of the PHE signal depends on the applied field amplitude means that the magnetization does not remain fully saturated during this rotation, but is probably distributed at the microscopic scale within an angular sector around the average drag angle. The higher the rotating field amplitude, the narrower this angular sector. In Fig. 5(b), the angular dependence of the sine of the drag angle derived from the PHE phase shift is plotted versus field amplitude. The fit of this variation with eq. 6 is quite good and yields $\boldsymbol{\beta = 1.67 \times 10^9\, rad/s}$.

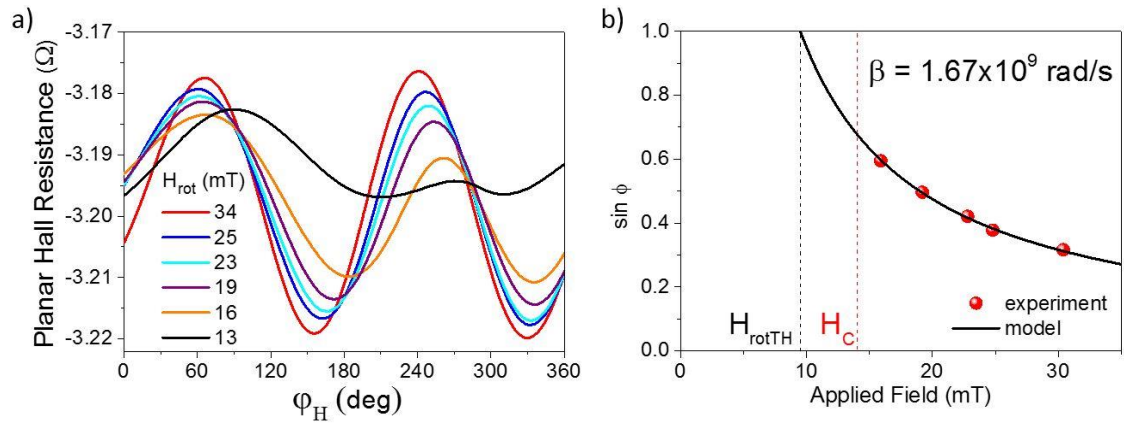

Fig.5 a) Planar Hall resistance under rotating field. b) Experimental angular shift (red points) fitted by eq. 6 (black line).

Moreover, the value of the coercive field of the sample (in red dashed line in Fig. 5b) is found to be different from the threshold field predicted by the fitting curve. Again, a reduction in the total magnetization during the motion can explain the higher value of the coercive field found experimentally with respect to the one found by the model.

In conclusion, it was shown in this first part of the paper that an isotropically coercive layer can be realized using a ferromagnetic/antiferromagnetic/ferromagnetic trilayer in which the antiferromagnetic layer is adjusted just below the onset of exchange bias. The isotropic coercivity in such system can be described by a dry friction torque introduced in the LLG equation. In the next section, we investigate by numerical simulation the possibility to manipulate the magnetization of this isotropically coercive layer by spin transfer torque to achieve a memristive function.

## III. PART 2: CURRENT-DRIVEN DYNAMICS UNDER DRY FRICTION

### A. Perpendicular polarizer STT-driven dynamics under DC current

As a further step towards the realization of a memristor based on the angular variation of the tunnel magnetoresistance of MTJs, the influence of the STT due to an additional perpendicular polarizer (P) on such isotropically coercive in-plane magnetized free layer was investigated by simulation. This perpendicular configuration has been widely studied for Spin Torque Nano Oscillators and for fast switching precessional MRAM ([29] – [34]) (Fig 6a). As described by Ebels et al. [29], under DC current, the influence of the STT from the perpendicular polarizer on the in-plane free layer is to slightly pull out-of-plane the magnetization of the free layer which then starts precessing around its demagnetizing field. In the steady state regime, the out-of-plane angle as well as the precession frequency linearly increase with the applied DC current up to a point where the magnetization gets saturated out of plane and stops precessing. Now if instead of DC voltage, successive voltage pulses are applied to the device, one can expect small in-plane step-by-step angular jumps of magnetization (depending on the amplitude and duration of the pulse) thus enabling the stabilization of intermediate levels of resistance between $R_{min}$ and $R_{max}$ depending on the sequence of voltage pulses. This is what we show below by numerical simulations.

From modelling point of view, eq. 3 is modified by including the torque $T = \gamma a_{//} V \mathbf{M} \times (\mathbf{M} \times \mathbf{p})$ with $\mathbf{p} = (0 \; 0 \; 1)$ the spin-polarization unitary vector, $a_{//}$ the coefficient in the Slonczewski term [35] given by $a_{//} = \frac{\hbar}{2e} \frac{\eta}{l\, M_S\, R \times A}$ where $\eta$ is the spin polarization, $l$ the layer thickness, $V$ the voltage across the tunnel barrier separating the perpendicular polarizer and the free layer, and $R \times A$ the resistance-area product of this tunnel barrier. The value used in this work is $a_{//} = 12\, mT/V$, corresponding to $\eta = 0.7$, $RxA = 10\, \Omega \cdot \mu m^2$, $l = 2nm$. When the dry friction is considered, two regimes can occur separated by a threshold voltage $V_{TH} = \beta/(\gamma a_{//})$ (Fig. 6b). For voltages lower than this threshold, the friction is stronger than the STT leading to an in-plane stable state of the magnetization. In contrast, above the voltage threshold, the free layer magnetization is slightly pulled out-of-plane and the precession starts, initially with a relatively low frequency, then increasing linearly with the voltage.

In this last case, in steady precession state, the precession frequency is given by:

$$\omega_M = -\frac{\gamma a_{//} V}{\alpha_G} - \frac{\beta}{\alpha_G sin(\theta_M)}\frac{\omega_M}{|\omega_M|} \qquad (7)$$

and the out-of-plane normalized component reads

$$cos(\theta_M) = \frac{a_{//} V}{\alpha_G M_S} + \frac{\beta}{\alpha_G \gamma M_S sin(\theta_M)}\frac{\omega_M}{|\omega_M|} \qquad (8)$$

where $\frac{\omega_M}{|\omega_M|}$ indicates that the precession occurs in opposite directions (clockwise vs anticlockwise) and the out-of-plane component of magnetization changes sign for opposite polarities of the voltage in eq.7 and eq.8.

As indicated by eq. 7 and eq. 8, the effect of dry friction shows up as a shift in both frequency and out-of-plane component as shown in Fig. 6c and 6d. The macrospin simulations confirm that for low values of the friction parameter, the well-known linear behavior reported in reference [29] is recovered (black lines in Fig. 6c and 6d). Note that differently from the case $\beta = 0$, the maximum value (corresponding to the case in which the magnetization is stable along the z-axis and no more dynamics occurs) is only reached asymptotically at infinite voltage.

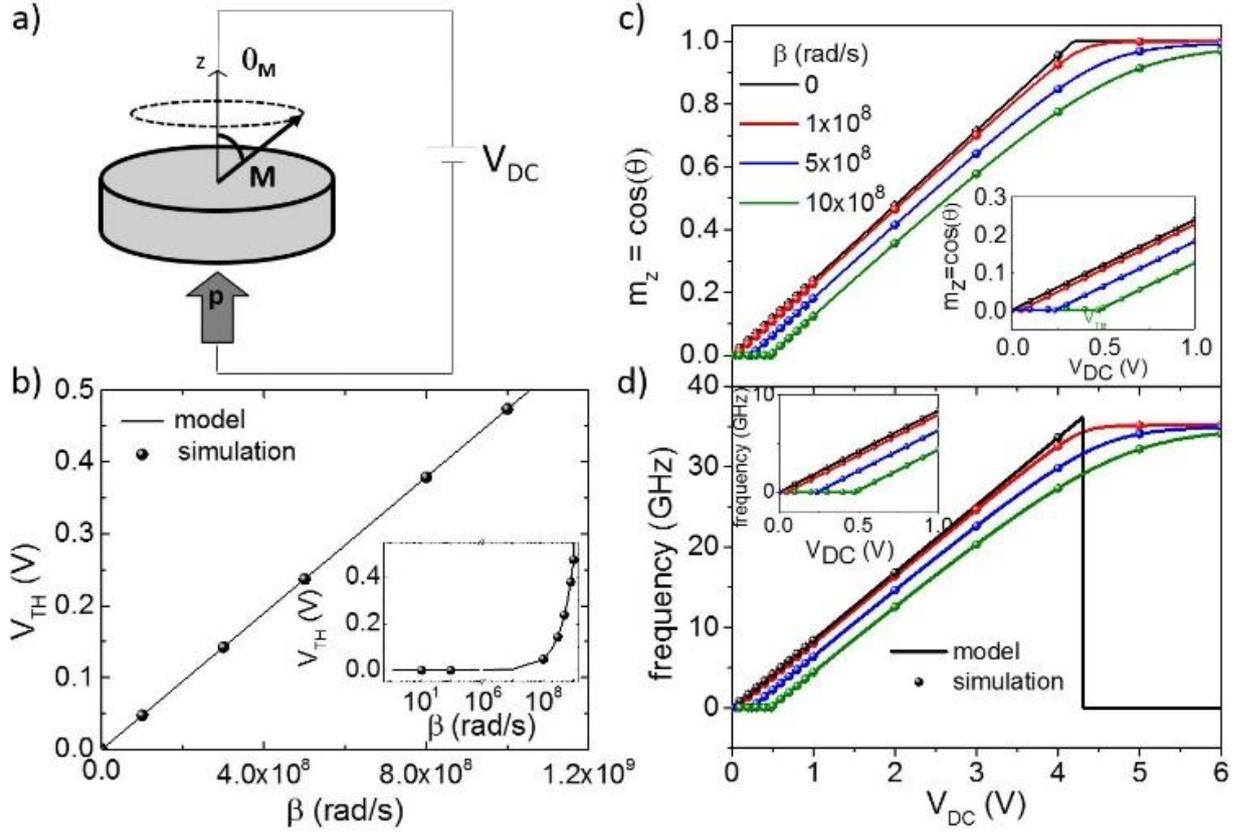

Fig.6 a) Schematic of the STT effect on a sample with a free layer **M** submitted to dry friction and a perpendicular polarizer **p**. b) Linear behavior of the threshold voltage with $\beta$ (in the inset the log scale). c) and d) free layer out-of-plane component and precession frequency versus dry friction amplitude. Insets: zoom around the voltage thresholds. The analytical model and the macrospin simulations are shown respectively in lines and dots for different values of $\beta$.

### B. In-plane field - DC current diagram with perpendicular polarizer

As explained in the general introduction, a transverse field can be used to limit the maximum excursion angle of the free layer magnetization in order to keep a fundamental property of a memristor: for the memristor resistance to be used as a synaptic weight, its resistance must vary monotonously for each current pulse polarity i.e. increase for one pulse polarity and decrease for the opposite pulse polarity ([1] and [2]). In the following, we shown that this can be achieved by applying a static transverse field of appropriate amplitude on the free layer. To start with, the combined effect of the STT due to an applied DC voltage and an in-plane static field is first considered by introducing both terms in $\boldsymbol{H_{eff}}$ and solving eq. 3. In the case where the field $H_{ext}$ is applied along the y-axis, the resulting effective field can be written

$$\boldsymbol{H_{eff}} = (Cm_y, \quad -Cm_x + H_{ext}, \quad -M_s m_z)$$

where $C = \frac{a_{//}}{\mu_0}V$.

In the numerical results shown in Fig. 7a, we can distinguish four regions: when the sum of the torques is lower than the dry friction (region 1), when one of the two is dominant (region 2 and 3) and when the two are competing (region 4).

In region 1, when the sum of the two torques is lower than the dry friction, magnetization motion cannot occur. The evident asymmetry in this region is due to the constructive or destructive competition of the two torques.

In region 2, the torque due to STT is prevailing leading to out-of-plane steady state precession of the magnetization. The effect of the increasing field is to tilt the plane containing the precessing magnetization trajectory.

In regions 3 and 4, the effect of STT acts as an in-plane discrete rotation of the magnetization that stays stable for all the pulse duration [29]. The static nature of the two torques can be analyzed as in the previous sections. In this case, looking for static $\frac{\partial \mathbf{M}}{\partial t} = 0$ and in-plane solutions ($\theta_M = \pi/2$), it can be found that the equilibrium condition becomes

$$|\gamma(H_{ext} m_x - C)| < \beta$$

where, in this particular case, $m_x = \sin(\varphi_H - \varphi_M)$. The final result writes

$$-\frac{\beta}{\gamma \mu_0 H_{ext}} + \frac{a_{//}}{\mu_0 H_{ext}} V < \sin(\varphi_H - \varphi_M) < \frac{\beta}{\gamma \mu_0 H_{ext}} + \frac{a_{//}}{\mu_0 H_{ext}} V \qquad (9)$$

The creation of a stability sector around the effective field, as in Fig. 2, is modified by the STT linearly proportional to the applied voltage. It is interesting to note that an asymmetry with respect to the direction of the applied field is induced by the voltage term as shown in Fig. 7b. In fact, depending on the polarity of the applied voltage, the scalar product of the two components of the torque (field and STT) is positive or negative. Thus, the amplitude of the cone becomes only dependent on the field amplitude and the effect of the applied voltage is explicit in the angular shift. While in region 3 the effect of the field is dominant, leading to a small sector of stability more or less centered along the field direction (small angular shift), in region 4, the angular shift becomes significant, forcing the magnetization to point towards the positive or negative x-axis direction depending on the polarity of the voltage.

Moreover, the in-plane stable states created in this way can be destabilized for a threshold value of the voltage $V_{TH}$ for which out-of-plane precession starts to occur (limit of the region 1-2 and 2-4). This threshold value is given by:

$$V_{TH} = \frac{\mu_0 H_{ext}}{a_{//}} - \frac{\beta}{a_{//}\gamma}$$

As shown in Fig. 7a, the analytical white lines are well fitting the oscillations limits obtained through macrospin simulations.

It is important to note that the results in Fig. 7a are obtained by applying first the field at t = 0 and then the voltage with a certain time delay (experimentally the field is supposed to be applied all the time). In this case, a first relaxation of the magnetization can occur towards the cone and only after the effect of the angular shift on the final magnetization state is considered.

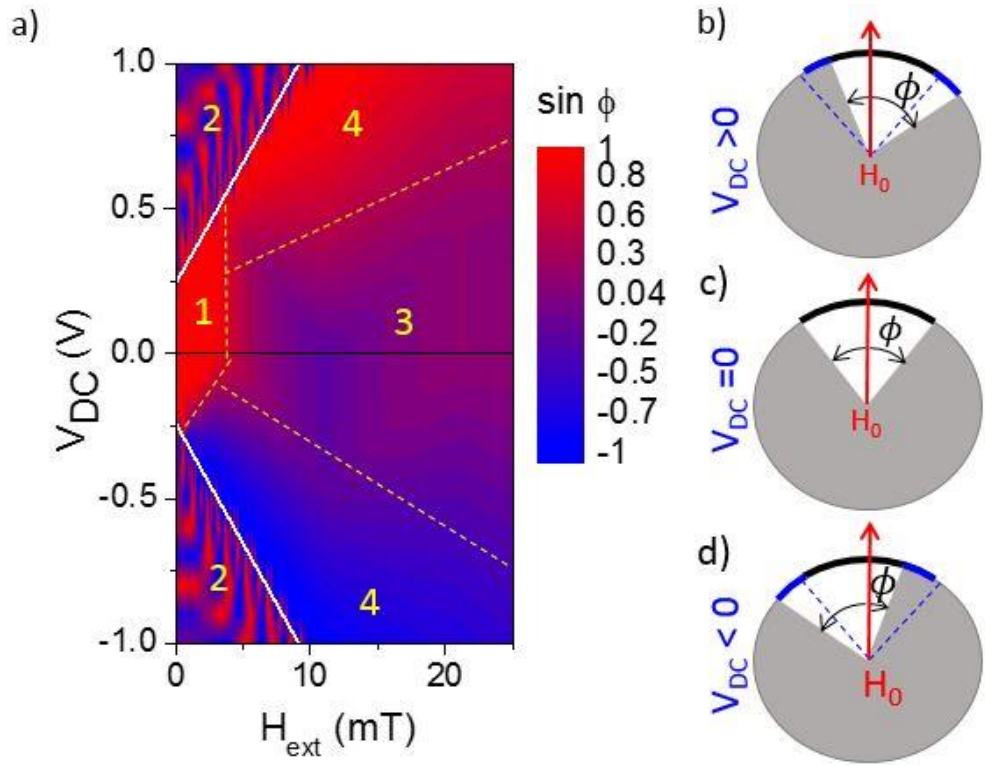

Fig. 7 a) Voltage/Field diagram for $\beta = 5 \times 10^8 rad/s$. The white lines are the analytical limit of the oscillating region. b) c) and d) Schematics of the combined effect of field and voltage on the system respectively for positive, zero and negative voltages.

Finally it is important to underline that the dependence of the in-plane stable states (region 3 and 4) on voltage cannot be exploited for the memristive characteristics. In fact, despite the linear behavior shown in eq. 9, the monotonicity of the rotation with respect to the duration of the pulse is not respected (there is not any time dependence in the equation). Moreover, the application of two consecutive identical pulses will not have any effect because the magnetization will rotate forth and back exactly by the same angle when the pulse is applied and when it is switched off. For these reasons, the only way to obtain the wanted memristive characteristics is by exploiting the out-of-plane oscillating regime of region 2, achieving small angle rotations through the use of pulses.

**C. Analyzer STT effect**

The full device with an out-of-plane polarizer (P), an in-plane free layer with dry friction (FL) and an in-plane analyzer (A), as shown in Fig. 1a, is described in the macrospin code to simulate the complete memristor characteristics. Two STT terms must then be considered originating on the one hand from the bottom perpendicular polarizer, as already discussed, and on the other hand from the top in-plane pinned analyzer.

Firstly, the effect of the STT term due to the analyzer is studied and compared to the one of the polarizer. An in-plane field / dc voltage diagram is simulated in order to find the out-of-plane oscillating region at different conditions: for each of the two STT contributions (separately), the in-plane field is applied parallel and perpendicular to the analyzer magnetization direction, for $\beta = 0 \ rad/s$ and for $\beta = 5x10^8 rad/s$. The two junctions are supposed to be identical, with a spin polarization of 70%. The results are shown in Fig. 8.

Starting from the analyzer STT contribution, it is clear that, for $\beta = 0 \ rad/s$ (Fig. 8a), the effect of the transverse field (red phase diagram) is to shift the threshold of oscillations towards higher voltage (higher than the simulated one) with respect to the case with field parallel to the magnetization (blue phase diagram). The same effect is obtained by adding the dry friction term to the system with field parallel (Fig. 8b in blue) where, in the voltage range simulated, it is not even possible to switch to the antiparallel configuration. We can conclude that the sum of both effects (dry friction and transverse field, as in the previous section and in red in Fig. 8b) is summing up in a relevant shift of the threshold voltage.

A similar effect occurs when the polarizer STT contribution is considered. In this case the direction of the field is not affecting the limits of the oscillating region because of an evident symmetry (Fig. 8c blue and red). Moreover, the effect of the dry friction, evident in a shift of the threshold (as discussed in section 2.1), is much smaller than the one observed for the only analyzer. This can be explained by considering that the initial velocities associated to the two effects are very different. In fact, the slow beginning of the precessing motion, in case of the in-plane analyzer, is easily stopped by the presence of the dry friction term, while in the case of polarizer contribution, the beginning of the precession is much faster.

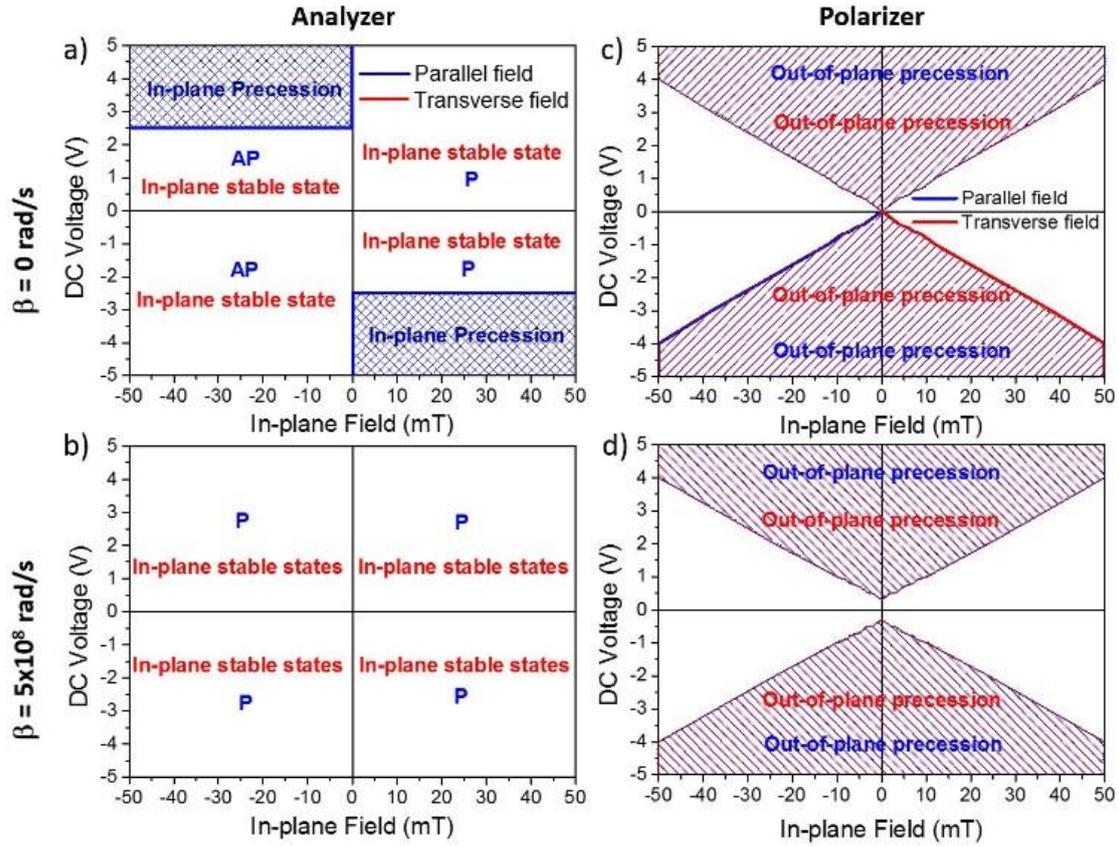

Fig. 8 a) Dynamical phase diagram for only STT from analyzer with $\beta = 0\ rad/s$ for in-plane field applied parallel (blue) and perpendicular (red) to the analyzer magnetization direction. b) Same diagram as in a) with $\beta = 5\text{x}10^8 rad/s$. c) Dynamical phase diagram for only STT from perpendicular polarizer with $\beta = 0\ rad/s$ for in-plane field applied parallel (blue) and perpendicular (red) to the analyzer magnetization direction. b) Same diagram as in c) with $\beta = 5\text{x}10^8 rad/s$.

### D. Simulation of the complete device

Finally, the dynamical phase diagram of the full device under DC current and trains of current pulses, considering the sum of the two STT contributions coming from the two junctions, is shown in Fig. 9. The diagram under DC current (Fig.9a) is actually similar to the one in Fig. 8d in which only the influence of the STT from the perpendicular polarizer was considered. Indeed, the effect of the STT from the in-plane analyzer in this field / voltage range is only to tilt the plane of trajectory of the precessing magnetization but it does not change the boundaries of the diagram. This is coherent with the results obtained in reference [36], where the effect of the analyzer is not to affect the out-of-plane precession region but only to add another region of in-plane oscillations.

The static field is applied in the perpendicular in-plane direction with respect to the top analyzer magnetization direction. It corresponds to the transverse field discussed in the previous section. Thus, if the magnetization goes from one edge to the other of the created sector of stability, the resistance will increase or decrease depending on the voltage pulse polarity (one of these edges is close to the parallel configuration of the top junction while the other is close to the antiparallel one). Note that the bottom junction is not giving any signal in terms of TMR since it remains in an invariant 90° configuration.

Fig. 9c and d show the free layer magnetization response to a series of positive and negative voltage pulses as in Fig. 8d respectively without and with a transverse in-plane field of 3.5mT (just above the threshold). This field allows the formation of a sector of stability (in white in Fig. 9d) that, as in section 1, is narrower than 180°. After a fast relaxation of the magnetization inside the sector of stability, a series of identical positive pulses of 100ps at 0.8V is applied. The pulse duration is adjusted to be a fraction of the half period of oscillations which itself depends on the voltage as given by eq.7. As a result, several intermediate values of resistance can be reached. When the magnetization gets close to the limit of the sector of stability (after 70ns in the plot of Fig. 9c), the voltage pulse pulls the magnetization out of it (in the grey region), but as soon as the voltage pulse ends, the magnetization relaxes back to within the sector of stability due to the torque from the transverse field (exactly as explained in Fig. 2c and d). This effect can be used to definitely limit the maximum magnetization excursion angle to a region in which the resistance variation is monotonous as required for a memristor (Fig. 1f) ([1] and [2]). The non-uniform rotation of the magnetization with identical pulses in Fig. 9d is due to the fact that while the spin transfer torque from the perpendicular polarizer is isotropic as is the dry friction torque, the torque due to the transverse field tends to attract the magnetization towards its direction with an amplitude proportional to the sine of the angle between magnetization and field. As a result, for each polarity of the voltage, there is a region in which the rotation angle is larger because the two torques (from STT and from field) favor the same direction of motion while in the other they oppose each other.

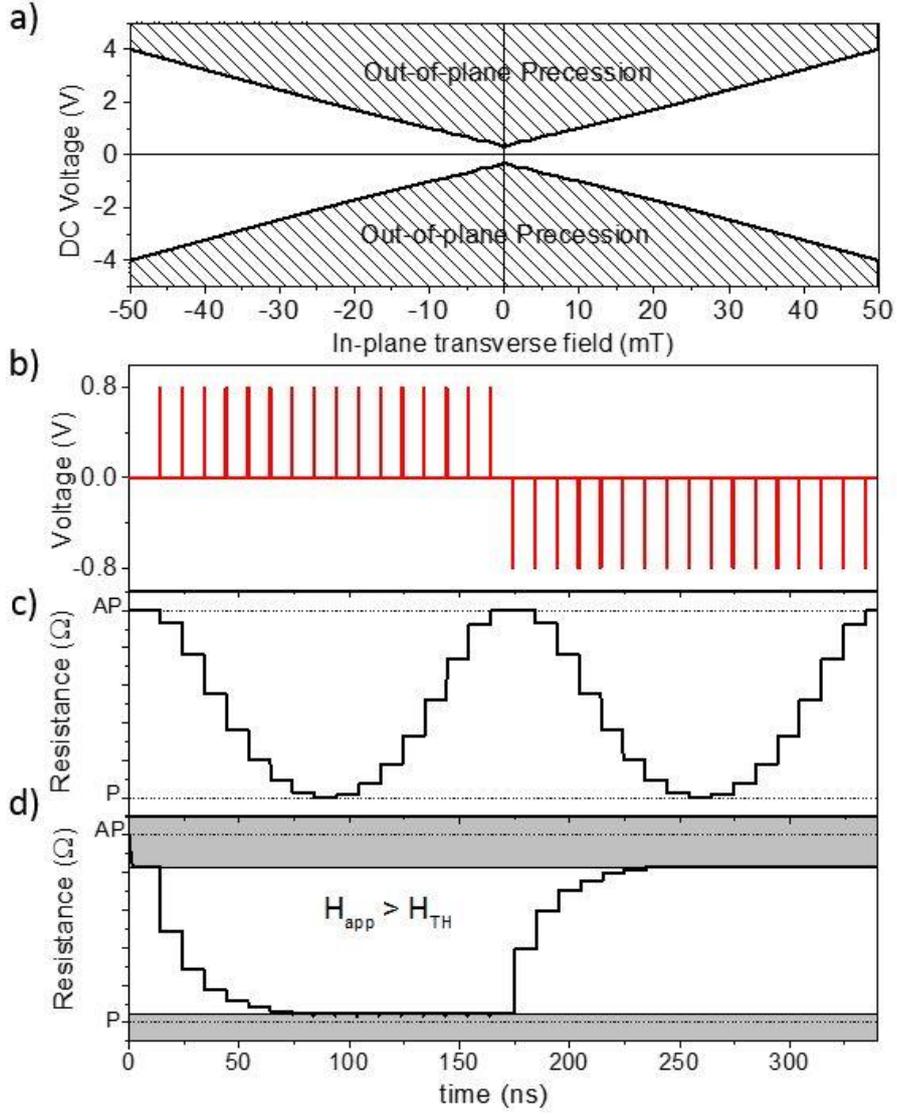

Fig.9 a) Field / DC voltage diagram of the device for $\beta = 5\mathrm{x}10^8 rad/s$. b) Train of pulses of 100ps and 0.8V. Raising time 50ps. c) Numerical simulation of the time evolution of the magnetization under a train of voltage pulses as in b), for $\beta = 5\mathrm{x}10^8 rad/s$. d) Numerical simulation of the time evolution of the magnetization under a train of voltage pulses as in b), for $\beta = 5\mathrm{x}10^8 rad/s$ and $H = 3.5$mT.

## IV. CONCLUSIONS

In the first part of the paper, we demonstrated that an isotropically coercive magnetic layer, needed for the realization of a spintronic memristor based on angular variation of TMR, can be realized by using exchange coupled F/AF/F trilayers in a regime where the AF is sufficiently thin for its spin lattice to be fully dragged during an evolution of the F magnetization. This form of dissipation can be described by inserting a dry friction term in the LLG equation. The fitting of the experimental values with the analytical expression allowed us to derive the value of the dry friction parameter $\beta = 1.67\mathrm{x}10^9 rad/s$. The resulting isotropic coercive field allows the magnetization to be stabilized along any in-plane direction.

In the second part, predictions on the effect of STT from a perpendicular polarizer on this system were made numerically and analytically. In this configuration (perpendicular polarizer/in-plane free layer with dry friction), the STT under DC current leads to an out-of-plane steady state precession whose frequency and out-of-plane component of magnetization vary linearly with voltage with a shift proportional to the dry friction coefficient. The effect of STT from perpendicular polarizer and in-plane reference layer combined with an in-plane static field allowed us to define a region wherein memristive behavior can be achieved. In this region, the application of short pulses, with duration corresponding to a small fraction of the oscillating period, allows a step by step in-plane rotation of the free layer magnetization limited between 0 and 180° thus realizing a memristive behavior.

**Acknowledgements**

This work was funded under the ERC Adv grant MAGICAL N° 669204.